# Enhanced Radar Imaging Using a Complex-valued Convolutional Neural Network

Jingkun Gao, Bin Deng, Yuliang Qin, Hongqiang Wang and Xiang Li

*Abstract*—Convolutional neural networks (CNN) have been successfully employed to tackle several remote sensing tasks such as image classification and show better performance than previous techniques. For the radar imaging community, a natural question is: Can CNN be introduced to radar imaging and enhance its performance? The presented letter gives an affirmative answer to this question. We firstly propose a processing framework by which a complex-valued CNN (CV-CNN) is used to enhance radar imaging. Then we introduce two modifications to the CV-CNN to adapt it to radar imaging tasks. Subsequently, the method to generate training data is shown and some implementation details are presented. Finally, simulations and experiments are carried out, and both results show the superiority of the proposed method on imaging quality and computational efficiency.

## I. INTRODUCTION

IMPROVING radar imaging quality using signal processing techniques under given hardware platforms is an appealing topic and has attracted much attention [1],[2]. In the recent decade, compressive sensing [3] has greatly inspired the research on sparsity-driven radar imaging techniques [4],[5]. Typically, the following linear model is adopted

$$y = Ax + n,\ y \in \mathbb{C}^{M\times 1}, n \in \mathbb{C}^{M\times 1}, x \in \mathbb{C}^{N\times 1}, A \in \mathbb{C}^{M\times N} \quad (1)$$

where $y$ is the echo signal, $n$ is the additive noise, $x$ represents the image to be estimated and $A$ is called the sensing matrix or the dictionary. According to (1), conventional imaging can be expressed as $\hat{x} = A^H y$, which can be achieved by back projection algorithm (BPA) or by Fast Fourier Transform (FFT). However, either methods suffer from limited resolution, high side-lobes and strong speckle. Sparsity-driven methods impose prior constrains to $x$ and the imaging problem is transformed into an optimization problem

$$\hat{x} = \arg\min_{x} \psi(x) \quad \text{s.t.} \quad \|y - Ax\|_2^2 < \sigma^2 \quad (2)$$

where $\psi(x)$ contains the prior knowledge on $x$, $\sigma^2$ represents the energy of $n$ which is always unknown. The most widely used prior is the $l_1$-norm constraint in the image domain, i.e. $\psi(x) = \|x\|_1$. The constraint is also usually imposed to the transformed domain, i.e. $\psi(x) = \|Tx\|_1$, where $T$ can be the wavelet transform, the derivative operator and so on. More complicated $\psi(x)$s which take the inner coherence of $x$ into consideration are also proposed [6]. It can be expected that one can obtain better $\hat{x}$ by designing more sophisticated $\psi(x)$ and developing more effective optimization algorithms.

Although sparsity-driven methods can improve imaging quality remarkably, it faces great challenges. Firstly, they are too time-consuming to achieve real time imaging. Different from conventional linear imaging process, the problem in (2) is nonlinear. Usually, a large amount of iterations are needed to converge to a reasonable solution. Secondly, their stability and robustness cannot easily be guaranteed. The imaging quality promotion obtained by solving (2) is based on accurate modeling. If $A$ is inaccurate or the constraint $\psi(x)$ is unsuitable, the results can be greatly degraded. Although several methods have been proposed for these problems [7], it is at the cost of heavier computational burdens.

In fact, the model in (1) and (2) is a general problem solving framework. In many signal processing fields, e.g. image processing or computer vision [8], this model is also widely used. CNN is famous mostly for its cutting-edge performance in image classification tasks [9]. Recently, more and more researchers have extended and applied CNN to regression-type problems such as superresolution or denoising [10],[11]. In [10], the authors showed the close relationship between regression-type CNN and sparsity-based methods. In [12], the authors pointed out the similarity between the iterative shrinkage methods and the feedforward process of CNN. Moreover, CNN can learn from and be adaptive to the training data, and its structure is highly parallel and free of iterations. As a result, CNN can outperform sparsity-based methods in both accuracy and efficiency.

On top of the above review, a reasonable question is whether CNN can be applied to radar imaging. In our opinion, several issues need to be addressed when applying CNN to radar imaging. For example, what is the overall processing framework? What is the input and output of the CNN? How can one obtain the training data? Is CNN more effective than recent imaging methods? In the following sections, we will introduce our solutions to these issues.

## II. METHODS

### A. Overall framework of CV-CNN-enhanced radar imaging

The proposed processing framework is shown in Fig. 1, the dashed and solid lines show the data flow in training and imaging/testing period respectively. The imaging model contains all configurations and parameters that imaging needs.

The proposed framework is actually an intuitive application of the idea of supervised learning to radar imaging tasks. If we regard the imaging processor as a black box, the input and output of this block are the echo data and the formed images respectively. In Fig. 1, we replace the conventional imaging processor with a CV-CNN. For conventional imaging processor, the algorithms are defined according to the developers' knowledge on imaging. Differently here, the imaging processor, i.e. the CV-CNN, is self-developed by training with given input and output examples. It can be seen that both training and imaging rely on the imaging model. Therefore, the premise of CV-CNN-enhanced radar imaging is to confirm the imaging model. Detailed discussions on the modifications on the CV-CNN and the method of training data generation will be given in the following two sections.

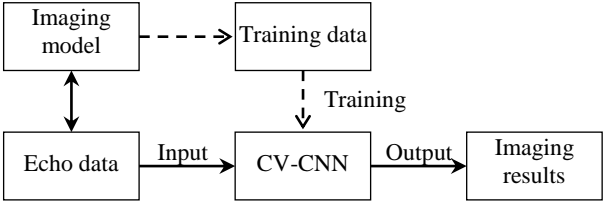

Fig. 1. Framework of CV-CNN-enhanced radar imaging

### B. Modifications on the CV-CNN

There were articles talking about CV-CNN in 1990s [13]. More recently, some publications rediscovered CV-CNN and recognized some of its advantages [14],[15]. Fig. 2 shows the basic neuron structure and connections in CV-CNN.

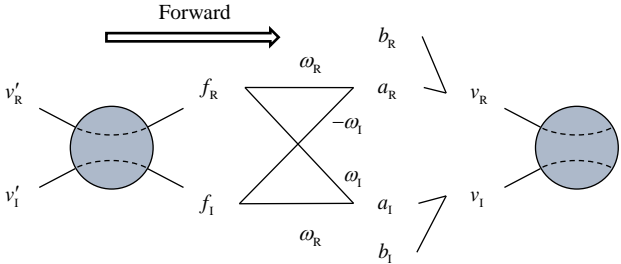

Fig. 2. Neuron and basic connection in CV-CNN

$v'$ and $v$ are the input of the neurons at lower and higher levels respectively, $f$ is the output of the neuron, $\omega$ is the weight, $b$ is the bias, and $v', f, \omega, a, b, v \in \mathbb{C}$. Subscript "R" and "I" represent the real and imaginary parts respectively. Suppose the cost function of CV-CNN is $E \in \mathbb{R}$ which will be defined latter. The training process, i.e. the complex-valued back propagation, can be summarized into following formulas [13],[15]

$$\begin{cases} \partial E/\partial \omega_R = \partial E/\partial v_R \cdot \partial v_R/\partial \omega_R + \partial E/\partial v_I \cdot \partial v_I/\partial \omega_R \\ \partial E/\partial \omega_I = \partial E/\partial v_R \cdot \partial v_R/\partial \omega_I + \partial E/\partial v_I \cdot \partial v_I/\partial \omega_I \end{cases} \quad (3)$$

$$\begin{cases} \partial E/\partial v'_R = \left(\partial E/\partial v_R \cdot \omega_R + \partial E/\partial v_I \cdot \omega_I\right) \cdot \partial f_R/\partial v'_R \\ \partial E/\partial v'_I = \left(-\partial E/\partial v_R \cdot \omega_I + \partial E/\partial v_I \cdot \omega_R\right) \cdot \partial f_I/\partial v'_I \end{cases} \quad (4)$$

To make the CV-CNN more efficient and suitable for radar imaging. We introduce two modifications: a new activation function and a new type of neuron for the output layer.

In [15], the conventional sigmoid function is extended to CV-CNN. However, many recent publications suggest that Rectified Linear Unit (ReLU) is a more effective activation. Therefore, we are extending ReLU to its complex-valued version in this letter. We define the complex-valued ReLU (cReLU) as follow

$$f = \text{cReLU}(v') = \max(v'_R, 0) + j\max(v'_I, 0) \quad (5)$$

and we can obtain

$$\partial f_R/\partial v'_R = \begin{cases} 1, & v'_R > 0 \\ 0, & v'_R \leq 0 \end{cases}, \partial f_I/\partial v'_I = \begin{cases} 1, & v'_I > 0 \\ 0, & v'_I \leq 0 \end{cases} \quad (6)$$

It can be seen that we simply apply the ReLU to the real and imaginary parts of the neuron input independently, which is effective and also critical to simplify the back propagation process. Similarly, a leaky version cReLU can also be defined according to (5).

In most radar imaging scenarios, people mainly care about the amplitude of the images. Also, focusing on the amplitude information of the image helps us to simplify some later issues such as training data generation. Consequently, we hope the outputs of our CV-CNN are real-valued. Therefore, a new type of neuron for the output layer is specifically defined as shown in Fig. 3.

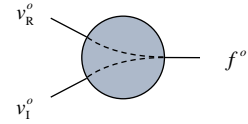

Fig. 3. Structure of the output layer neuron

According to Fig. 3, we define the corresponding output layer activation function as

$$f^o = \left|v^o_R + jv^o_I\right| = \sqrt{\left(v^o_R\right)^2 + \left(v^o_I\right)^2} \quad (7)$$

where the superscript "$o$" indicates the output layer, and $f^o \in \mathbb{R}$. We can name this function as "Abs". Then the definition of cost function can be given

$$E = \frac{1}{2}\sum\left(O - f^o\right)^2 \quad (8)$$

where $\sum\cdot$ is applied to all the output layer neurons, $O \in \mathbb{R}$ is the target function which will be defined in next section. The error term of the output layer are

$$\partial f^o/\partial v^o_R = v^o_R/f^o, \partial f^o/\partial v^o_I = v^o_I/f^o \quad (9)$$

Given all the above definitions and modifications, we can train the proposed CV-CNN accordingly.

### C. Generation of the training data

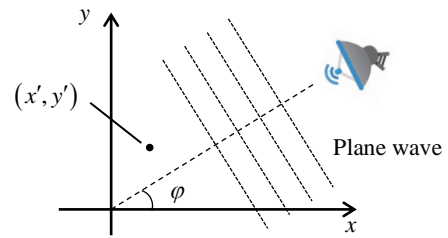

Fig. 4. Turntable imaging model

As aforementioned, the premise of training data generation is the knowledge of the imaging model. In this letter, we will take the classical turntable model shown in Fig. 4 as an example. For other typical imaging models such as SAR/ISAR, the ideas to generate training data are similar.

According to Fig. 1, the input and output of the CV-CNN are radar echoes and expected images. For real targets, the echoes are determined by the electromagnetic scattering equation

$$\boldsymbol{E}^s(\boldsymbol{r}) = \frac{jk\exp(jkR_0)}{4\pi R_0}\hat{k}_s \times \int_S \left(\hat{n}\times\boldsymbol{E}(\boldsymbol{r}') - \eta_0\hat{k}_s\times(\hat{n}\times\boldsymbol{H}(\boldsymbol{r}'))\right)\exp\left(-jk(\hat{k}_s - \hat{k}_i)\cdot\boldsymbol{r}'\right)d\boldsymbol{r}' \quad (10)$$

where $\boldsymbol{E}^s, \boldsymbol{E}, \boldsymbol{H}$ represent the scattered wave, the total electric and magnetic fields on target's surface respectively. $k$ is the wavenumber, $R_0$ is the distance from target to the observation point, $\hat{k}_s, \hat{k}_i$ are the unit vectors of the scattering and incidence direction, $S$ stands for the target's surface, $\hat{n}$ is the unit normal, $\eta_0$ is the free-space wave impedance. The echo is a complicated function of observation angle, polarization, frequency, and there is no explicit expression of the expected image function. To train CV-CNN, input and output must be exactly defined. Consequently, necessary simplification is applied to (10). Concretely, the ideal point scattering model is used and the scattering is confined within 2D space. Neglecting the attenuation and the propagation phase term, we can obtain the echo under turntable model as

$$E^s(k,\varphi) = \sum_{x',y'} o(x',y')\cdot\exp\left(-2jk(x'\cos\varphi + y'\sin\varphi)\right) \quad (11)$$

where $o(x',y')\in\mathbb{C}$ represents the target's scattering coefficients. Then we can explicitly defined the expected images as

$$O(x,y) = p(x,y) * |o(x,y)| \quad (12)$$

where $*$ stands for convolution, $O(x,y)\in\mathbb{R}$ is the same function in (8) and $p(x,y)$ is the ideal point spread function (PSF). We define $p(x,y)$ as

$$p(x,y) = \exp\left(-x^2/\sigma_x^2 - y^2/\sigma_y^2\right) \quad (13)$$

where $\sigma_x, \sigma_y$ control the width of $p(x,y)$. According to the $-3\,\text{dB}$ definition, we can deduce the resolution in these two directions are $1.18\sigma_x, 1.18\sigma_y$ respectively.

While generating training data, we first randomly generate hundreds of coordinates $(x',y')$ within the given area using uniform distribution, and the scattering coefficients are generated using standard complex Gaussian distribution, i.e. $o(x',y') \sim N(0,1) + jN(0,1)$. Then training data at the input and output ports of the CV-CNN are obtained by (11) and (12) respectively. We want to clarify that the randomly generated coordinates $(x',y')$ are not discretized or confined to integer image cells. They are made arbitrary continuous real numbers so as to imitate the real-world cases upmost.

## III. RESULTS

### A. Network structure and implementation details

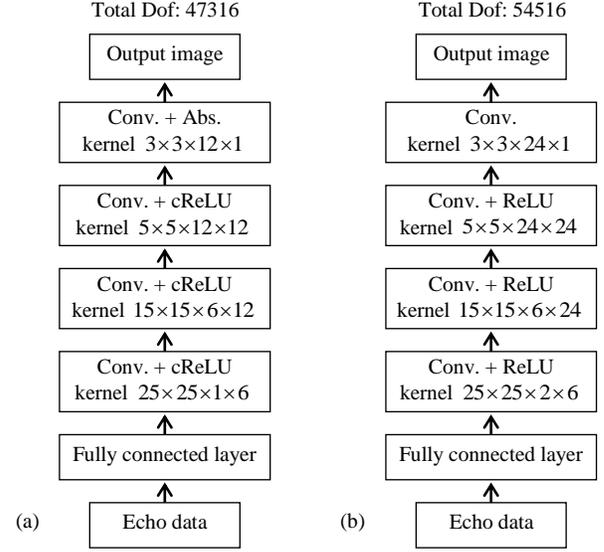

Fig. 5. Network structures of (a) proposed CV-CNN and (b) the counterpart RV-CNN

The fully connected layer in our network is achieved by $\hat{\boldsymbol{x}} = \boldsymbol{A}^H\boldsymbol{y}$. For large scale problems, the number of elements in $\boldsymbol{A}$ can approach the order of $10^9$. Therefore, FFT is employed to realize $\hat{\boldsymbol{x}} = \boldsymbol{A}^H\boldsymbol{y}$ fast and implicitly. For general imaging cases, implicit fast implementation of $\boldsymbol{A}$ and $\boldsymbol{A}^H$ also exist [16]. As a result, the weights of the fully connected layer do not exist explicitly, and they are also excluded in the training phase. For comparison, a similar RV-CNN is also designed. The real and imaginary parts of the data are treated as two independent channels. The total degree of freedom (Dof) of the RV-CNN is designed to be higher than that of the CV-CNN. We can see that CNN-based imaging is actually a more general imaging approach and conventional algorithms can be treated as its special cases. For example, if we neglect the nonlinear activation function of the fully connected layer and all the following conv. layers and output layer, our network then degrade to the BPA which is a linear transformation.

Our recent implementation is based on MATLAB. Momentum stochastic gradient descent is used and the momentum is 0.9. Weight decay technique is also employed and the coefficient is 0.001. The learning rate is $3\times10^{-5}$ for the first three layers, and $1\times10^{-5}$ for the last layer. The batch size is 50, and 50,000 examples are generated for training. Training lasts for 5 epochs. One NVIDIA TITAN Xp card is used, and the training takes approximately 16 hours.

For sparsity-driven imaging, we take SPGL1 [17] as a representative to compete with the proposed method. There are three reasons that make SPGL1 a good choice. 1) It supports implicit implementation of $\boldsymbol{A}$. For large scale problems, explicit $\boldsymbol{A}$ will lead to unacceptable memory cost and block the computation. 2) It is compatible with complex-valued problems. Complex-valued data must be transformed to real ones to comply with real-valued algorithms. This doubles the size of the matrices and vectors, or it sometimes may lead to a wrong solution. 3) High executive efficiency. We have tested several algorithms and found that SPGL1 is relatively fast especially for large scale problems.

*B. Numerical simulations*

TABLE I
IMAGING PARAMETERS

| Probing frequency | 213.6 GHz ~ 226.4 GHz |
|---|---|
| Rotation angle $\varphi$ | -1.68 °~1.67 ° |
| Number of frequency samplings | 500 |
| Number of angle samplings | 300 |
| Region of imaging | 0.7×0.7 m$^2$ |
| Number of pixels in the image | 236×236 |

From Table 1, it is easy to calculate that the range and azimuth resolutions are 1.17 cm and 1.15 cm respectively. Both $\sigma_x$, $\sigma_y$ are set to be 0.4 cm which suggests a 0.47 cm resolution, so the expected superresolution ratio is about 2.5. According to [18], imaging can be done by 2D IFFT for this small rotation angle. In fact, the linear operator $A$ and $A^H$ in this letter are implicitly implemented by 2D IFFT and 2D FFT. It should be claimed that these fast implementations introduce some errors into $A$ and $A^H$ since FFT-based imaging actually approximates the fan-shaped spectral domain support area into a rectangular one. Fig. 6 shows the imaging results of the "NUDT" made up of several unit ideal point scatters. The groundtruth image generated by (12) is shown in Fig. 6(e). Also, we want to clarify that the coordinates of these point scatters are continuous and are not placed at the center of the image cells.

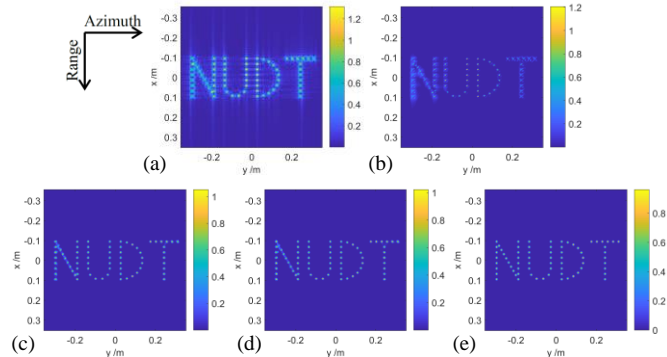

Fig. 6. Imaging results of "NUDT" by (a) FFT (b) SPGL1 (c) RV-CNN (d) CV-CNN, and (e) groundtruth image.

In Fig. 6(a), we can see that the image is of high side-lobes and relatively low resolution. We notice that the points located around the center are focused better than those close to the borders. In [19], this phenomenon is carefully analyzed, and it is caused by the errors introduced into $A$ we mentioned above. In Fig. 6(b), similar phenomenon can be observed since the errors in $A$ impact the solving of SPGL1. Compared with Fig. 6(a)(b), (c)(d) achieve higher quality, and they are also more similar to the groundtruth image in (e). For RV-CNN and CV-CNN, their recent fully connected layers are implicitly implemented by 2D IFFT. However, the networks can automatically fix these errors in $A^H$ since the following convolutional layers are adaptive. As a result, RV-CNN and CV-CNN are more robust to modeling errors. Quantitatively, the root mean square errors (RMSE) are calculated as performance index. The RMSEs of different methods under different SNRs are listed in Table 2. For each SNR level and each method, RMSE is the average of 100 times experiments' results. The testing data are generated in the same way with the training data. The time needs of different methods are the mean value of all the experiments for each method. For the time consumption of CNN-based imaging, it contains two parts. Firstly, the fully connected layer which is implicitly achieved by 2D-FFT is implemented on an Intel i3-4150 CPU. Then the data are transferred to the GPU where the following convolutions are calculated.

TABLE II
COMPARISON ON RMSES AND TIME NEEDS FOR DIFFERENT METHODS

| Methods | RMSE, -10dB | RMSE, -5dB | RMSE, 0dB | RMSE, 5dB | RMSE, 10dB | Time needs |
|---|---|---|---|---|---|---|
| FFT | 0.1987 | 0.1778 | 0.1705 | 0.1682 | 0.1675 | **0.042 s** |
| SPGL1 | 0.0568 | 0.0560 | 0.0568 | 0.0574 | 0.0683 | 27.66 s |
| RV-CNN | 0.0456 | 0.0308 | 0.0262 | 0.0251 | 0.0248 | 0.083 s |
| **CV-CNN** | **0.0434** | **0.0289** | **0.0255** | **0.0247** | **0.0245** | 0.071 s |

For all SNR levels, CV-CNN performs the best. For SPGL1, the SNR is imported into the algorithm as a known parameter which is not the case for real applications. Moreover, it can be seen that the RMSEs of SPGL1 is not monotonic. We think this is mainly caused by two reasons. 1) The algorithm is not quite stable since errors exist in $A$. 2) The groundtruth produced by (13) does not encourage image's sparsity too much. For RV-CNN, its performance is weaker than CV-CNN although there are 7200 more parameters. We think this is because the structure shown in Fig. 2 is essentially more efficient for modeling complex-valued problems. For the time needs, CV-CNN is closely after the FFT-based method. As the fully connected layer of CV-CNN is achieved by FTT and its time need can no way better than FFT-based method recently. Compared with SPGL1, the imaging process of CV-CNN is just the feedforward process of the network which can be easily parallelized and greatly accelerated by GPU. Therefore, CV-CNN costs much less time than SPGL1 and can meet the requirements for most real time imaging scenarios. The speedup of the proposed algorithm is not solely due to the utilization of the GPU. More importantly, different from traditional iterative regularization algorithms, our CNN-based algorithm relies on much simpler convolution operations which can be easily parallelized. Therefore, we believe the data listed in Table 2 are valuable references for practical applications.

*C. Laboratory results*

The experiment scenario is shown in Fig. 7 and an airplane model is used as the target. Other imaging parameters are the same with those listed in Table 1. In Fig. 8, images are displayed in log-magnitude and the dynamic range is 35 dB. The time needs are similar with those listed in Table 2 since parameters are the same. Unsurprisingly, the resolution of FFT-based image is relatively low while the side-lobes are extremely high. All other methods can enhance the resolution and suppress the side-lobes while their visual quality differ from each other apparently. Since it is hard to define groundtruth for real targets, quantitative comparisons can hardly be conducted. Moreover, the quality assessment of radar images is still an open question. Nevertheless, recent results have shown the superiority of the proposed method on robustness and efficiency.

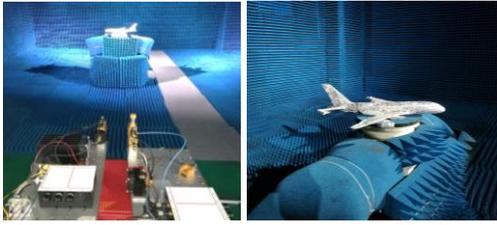

Fig. 7. Experiment scenario

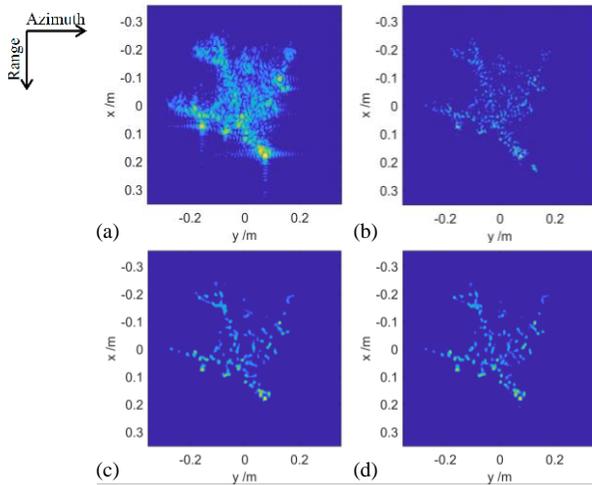

Fig. 8. Laboratory imaging results by different methods (a) 2D-FFT (b) SPGL1 (c) RV-CNN (d) CV-CNN.

In addition, we think CNN-enhanced radar imaging can be further improved in several aspects. The proposed CV-CNN is trained under ideal point scattering assumption. This indicates that current CV-CNN is actually point-feature-enhancement oriented. By using more accurate scattering models, enhancement of more potential features of the target can be expected. Also, we recently assume all needed imaging parameters are known. In real world scenarios, due to the uncertainty of movements, some parameters must be estimated. Whether CNN can be designed to achieve automatic focusing is also an appealing direction.

## IV. CONCLUSION

A novel CV-CNN-enhanced radar imaging method is proposed. The processing framework is firstly presented. The cReLU activation is proposed, and for the neurons in the output layer, the "Abs" activation is used. Then the method for generating training data is introduced. Different from iterative-based imaging methods, the CNN-based imaging process is just the feedforward process of the network. This makes it much faster and possible to achieve real time enhanced imaging. CNN-based method is also more robust since it is adaptive to the training data. Compared with its real-valued counterpart, CV-CNN achieves better performance with less parameters. We believe radar imaging will benefit more from deep learning techniques such as CNN in the future.

REFERENCES

[1] L. Zhao, L. Wang, L. Yang, A. M. Zoubir and G. Bi, "The Race to Improve Radar Imagery: An overview of recent progress in statistical sparsity-based techniques," *IEEE Signal Processing Magazine*, vol.33, no.6, pp. 85-102, 2016.
[2] S. Zhang, Y. Liu and X. Li, "Autofocusing for Sparse Aperture ISAR Imaging Based on Joint Constraint of Sparsity and Minimum Entropy," *IEEE Journal of Selected Topics in Applied Earth Observations & Remote Sensing*, vol.10, no.3, pp. 998-1011, 2017.
[3] D. L. Donoho, "Compressed sensing," *IEEE Transactions on Information Theory*, vol.52, no.4, pp. 1289-1306, 2006.
[4] G. Xu, M. Xing, L. Zhang, Y. Liu and Y. Li, "Bayesian Inverse Synthetic Aperture Radar Imaging," *IEEE Geoscience & Remote Sensing Letters*, vol.8, no.6, pp. 1150-1154, 2011.
[5] M. Cetin, I. Stojanovic, O. Onhon and K. Varshney, "Sparsity-Driven Synthetic Aperture Radar Imaging: Reconstruction, autofocusing, moving targets, and compressed sensing," *IEEE Signal Processing Magazine*, vol.31, no.4, pp. 27-40, 2014.
[6] L. Wang, L. Zhao, G. Bi and C. Wan, "Sparse Representation-Based ISAR Imaging Using Markov Random Fields," *IEEE Journal of Selected Topics in Applied Earth Observations & Remote Sensing*, vol.8, no.8, pp. 3941-3953, 2015.
[7] X. Zhou, H. Wang, Y. Cheng and Y. Qin, "Off-Grid Radar Coincidence Imaging Based on Variational Sparse Bayesian Learning," *Mathematical Problems in Engineering*, vol.2016, pp. 1-12, 2016.
[8] M. Elad, M. A. T. Figueiredo and Y. Ma, "On the Role of Sparse and Redundant Representations in Image Processing," *Proceedings of the IEEE*, vol.98, no.6, pp. 972-982, 2010.
[9] A. Krizhevsky, I. Sutskever and G. E. Hinton, "ImageNet classification with deep convolutional neural networks," *Communications of the Acm*, vol.60, no.2, pp. 2012, 2012.
[10] C. Dong, C. L. Chen, K. He and X. Tang, "Image Super-Resolution Using Deep Convolutional Networks," *IEEE Transactions on Pattern Analysis & Machine Intelligence*, vol.38, no.2, pp. 295-307, 2016.
[11] K. Zhang, W. Zuo, Y. Chen, D. Meng and L. Zhang, "Beyond a Gaussian Denoiser: Residual Learning of Deep CNN for Image Denoising.," *IEEE Transactions on Image Processing*, vol.26, no.7, pp. 3142-3155, 2017.
[12] K. H. Jin, M. T. Mccann, E. Froustey and M. Unser, "Deep Convolutional Neural Network for Inverse Problems in Imaging," *IEEE Transactions on Image Processing*, vol.26, no.9, pp. 4509-4522, 2017.
[13] G. M. Georgiou and C. Koutsougeras, "Complex domain backpropagation," *IEEE Transactions on Circuits & Systems II Analog & Digital Signal Processing*, vol.39, no.5, pp. 330-334, 1992.
[14] C. Trabelsi, O. Bilaniuk, Y. Zhang, D. Serdyuk, S. Subramanian*, et al.*, "Deep Complex Networks", *arXiv perpringt arXiv:1705.09792*, 2017.
[15] Z. Zhang, H. Wang, F. Xu and Y. Q. Jin, "Complex-Valued Convolutional Neural Network and Its Application in Polarimetric SAR Image Classification," *IEEE Transactions on Geoscience & Remote Sensing*, vol.55, no.12, pp. 7177-7188, 2017.
[16] J. Fang, Z. Xu, B. Zhang, W. Hong and Y. Wu, "Fast Compressed Sensing SAR Imaging Based on Approximated Observation," IEEE Journal of Selected Topics in Applied Earth Observation & Remote Sensing, vol.7, no.1, pp. 352-363, 2014.
[17] V. D. B. Ewout and M. P. Friedlander, "Probing the Pareto Frontier for Basis Pursuit Solutions," *Siam Journal on Scientific Computing*, vol.31, no.2, pp. 890-912, 2008.
[18] J. Gao, Y. Qin, B. Deng, H. Wang, J. Li and X. Li, "Terahertz Wide-Angle Imaging and Analysis on Plane-wave Criteria Based on Inverse Synthetic Aperture Techniques," *Journal of Infrared Millimeter & Terahertz Waves*, vol.37, no.4, pp. 373-393, 2016.
[19] J. Gao, Z. Cui, B. Cheng, Y. Qin, X. Deng*, et al.*, "Fast Three-Dimensional Image Reconstruction of a Standoff Screening System in the Terahertz Regime," *IEEE Transactions on Terahertz Science & Technology*, vol.8, no.1, pp. 38-51, 2018.